\documentclass[aip,rsi,reprint,amsmath,amssymb,superscriptaddress,floatfix]{revtex4-1}

 \usepackage[T1]{fontenc}
 \usepackage{amsmath}
\usepackage{tabularx}
 \usepackage{graphicx}
\usepackage{multirow}
 \usepackage{SIunits}
 \usepackage{ulem}
 \usepackage{braket}
 \usepackage{color}
 \usepackage[draft]{fixme}
\usepackage{amssymb}
\usepackage{dcolumn}
\usepackage{bm}
\usepackage{hhline}
\usepackage{float}
\usepackage{threeparttable}
\newcolumntype{P}[1]{>{\centering\arraybackslash}p{#1}}

\DeclareMathOperator*{\argmin}{arg\,min}
\DeclareMathOperator*{\argmax}{arg\,max}

\begin{document}

\title{Cryogenic photoluminescence imaging system for nanoscale positioning of single quantum emitters}

\date{\today}

\author{Jin Liu}
\email{liujin23@mail.sysu.edu.cn}
\affiliation{Center for Nanoscale Science and Technology, National Institute of Standards and Technology, Gaithersburg, MD 20899, USA}
\affiliation{Maryland NanoCenter, University of Maryland, College Park, MD 20742, USA}
\affiliation{School of Physics, Sun-Yat Sen University, Guangzhou, 510275, China}

\author{Marcelo I. Davan\c co}
\email{marcelo.davanco@nist.gov}
\affiliation{Center for Nanoscale Science and Technology, National Institute of Standards and Technology, Gaithersburg, MD 20899, USA}

\author{Luca Sapienza}
\affiliation{Center for Nanoscale Science and Technology, National Institute of Standards and Technology, Gaithersburg, MD 20899, USA}
\affiliation{Department of Physics and Astronomy, University of Southampton, Southampton SO17 1BJ, UK}

\author{Kumarasiri Konthasinghe}
\affiliation{Department of Physics, University of South Florida, Tampa, Florida 33620, USA}

\author{Jos\'e Vin\'icius De Miranda Cardoso}
\affiliation{Center for Nanoscale Science and Technology, National
Institute of Standards and Technology, Gaithersburg, MD 20899,
USA}
\affiliation{Federal University of Campina Grande, Campina Grande, Para\'iba, Brazil}

\author{Jin Dong Song}
\affiliation{Center for Opto-Electronic Materials and Devices Research, Korea Institute of Science and Technology, Seoul 136-791, South Korea}

\author{Antonio Badolato}
\affiliation{Department of Physics and Astronomy, University of Rochester, Rochester, NY 14627, USA}

\author{Kartik Srinivasan}
\email{kartik.srinivasan@nist.gov}
\affiliation{Center for Nanoscale Science and Technology, National Institute of Standards and Technology, Gaithersburg, MD 20899, USA}

\begin{abstract}
We report a photoluminescence imaging system for locating single quantum emitters with respect to alignment features.  Samples are interrogated in a 4~K closed-cycle cryostat by a high numerical aperture (NA=0.9, 100$\times$ magnification) objective that sits within the cryostat, enabling high efficiency collection of emitted photons without image distortions due to the cryostat windows.  The locations of single InAs/GaAs quantum dots within a $>50$~$\mu$m~$\times$~50~$\mu$m field of view are determined with $\approx4.5$~nm uncertainty (one standard deviation) in a 1~s long acquisition. The uncertainty is determined through a combination of a maximum likelihood estimate for localizing the quantum dot emission, and a cross-correlation method for determining the alignment mark center. This location technique can be an important step in the high-throughput creation of nanophotonic devices that rely upon the interaction of highly confined optical modes with single quantum emitters.
\end{abstract}

 \maketitle

 \section{Introduction}
 \label{sec:Intro}

 Single solid-state quantum emitters such as color centers in crystals and epitaxial quantum dots (QDs) are being explored for a variety of purposes in photonic quantum information technology, such as triggered single photon generation.  Although some techniques exist for the precise location of these emitters at the stage of their formation, such as site-controlled QD growth~\cite{baier_high_2004,ref:Schneider_site_controlled} or implantation through nanoapertures~\cite{ref:awschalom_NV_implantation}, much work continues to be focused on randomly positioned structures, such as naturally occurring impurity centers or self-assembled QDs.  In part, this is due to the optical quality of such structures.  For example, QDs produced by Stranski-Krastanov self-assembled growth~\cite{marzin_photoluminescence_1994,petroff_epitaxially_2001} have been used in single-photon sources exhibiting essentially transform-limited emission linewidths~\cite{kuhlmann_transform-limited_2015,ding_-demand_2016,somaschi_optimal_QD_sources}, which has not yet been shown for site-controlled QDs.  As a result, methods to locate the spatial position of quantum emitters with respect to alignment features become necessary in order to fabricate nanophotonic structures, such as gratings, waveguides, and microcavities~\cite{lodahl_interfacing_2015}, in which the quantum emitters are optimally aligned with respect to the confined optical modes.  In such devices, offsets in the position of the quantum emitters of $\gtrsim$~50~nm can result in a significant degradation in performance.

 Of the various techniques used to locate single quantum emitters such as QDs, including atomic force microscopy~\cite{hennessy_quantum_2007}, scanning confocal microscopy~\cite{lee_registration_2006,dousse_controlled_2008, thon_strong_2009}, scanning electron microscopy~\cite{badolato_deterministic_2005,kuruma_position_2016},and cathodoluminescence~\cite{gschrey_highly_2015}, photoluminescence imaging~\cite{kojima_accurate_2013,sapienza_nanoscale_2015} is particularly appealing due to the ability to use wide-field illumination and multiplexed detection on a sensitive camera to interrogate a large sample area in a relatively small amount of time.  In Ref.~\onlinecite{sapienza_nanoscale_2015}, we reported on the development of a photoluminescence imaging technique for locating single QDs with an uncertainty of $<30$~nm with respect to metallic alignment features ($<10$~nm when using a solid immersion lens), and demonstrated how this approach could be used in the creation of bright and pure triggered single-photon sources.  Here, we describe the performance of a second-generation setup that significantly improves upon our previous work in several ways. By working with a high numerical aperture objective housed within the closed-cycle cryostat used to interrogate the devices, we improve the collected photon flux, eliminate imaging distortion due to the cryostat windows, and ensure common mode vibration between the sample and objective.  We use a more sophisticated and automated data analysis method for determining the QD position, consisting of a maximum likelihood estimate~\cite{mortensen_optimized_2010} and cross-correlation approach~\cite{anderson_sub-pixel_2004} to localize the QD and alignment marks, respectively.  Taken together, these improvements result in a that is mean positioning uncertainty (one standard deviation value) $\approx4.5$~nm for an image acquisition time of 1~s, which represents more than a factor of 6$\times$ improvement in uncertainty and 100$\times$ reduction in acquisition time in comparison to our previous work.

 \section{Experimental Setup and Imaging Results}
 \label{sec:Main_Expts}

We interrogate samples grown by molecular beam epitaxy and consisting of a layer of low-density InAs QDs embedded at the midpoint of a GaAs layer (typically 190~nm thick), which in turn is grown on a $>1$~$\mu$m Al$_{0.65}$Ga$_{0.35}$As layer (used as a sacrificial layer in the fabrication of suspended devices like photonic crystal membranes).  We fabricate an array of alignment marks (typically crosses with a separation of 50~$\mu$m) on top of the samples through electron-beam lithography, Cr/Au deposition, and resist lift-off.  The samples are placed in a closed-cycle cryostat with top optical access and cooled to a temperature below 10~K. The cryostat also houses within it a 100$\times$ magnification, 0.9 numerical aperture (NA) objective lens that sits above the top of the radiation shield and below the top window that maintains the cryostat vacuum with respect to the lab environment.

\begin{figure}[t]
  \includegraphics[width=\linewidth]{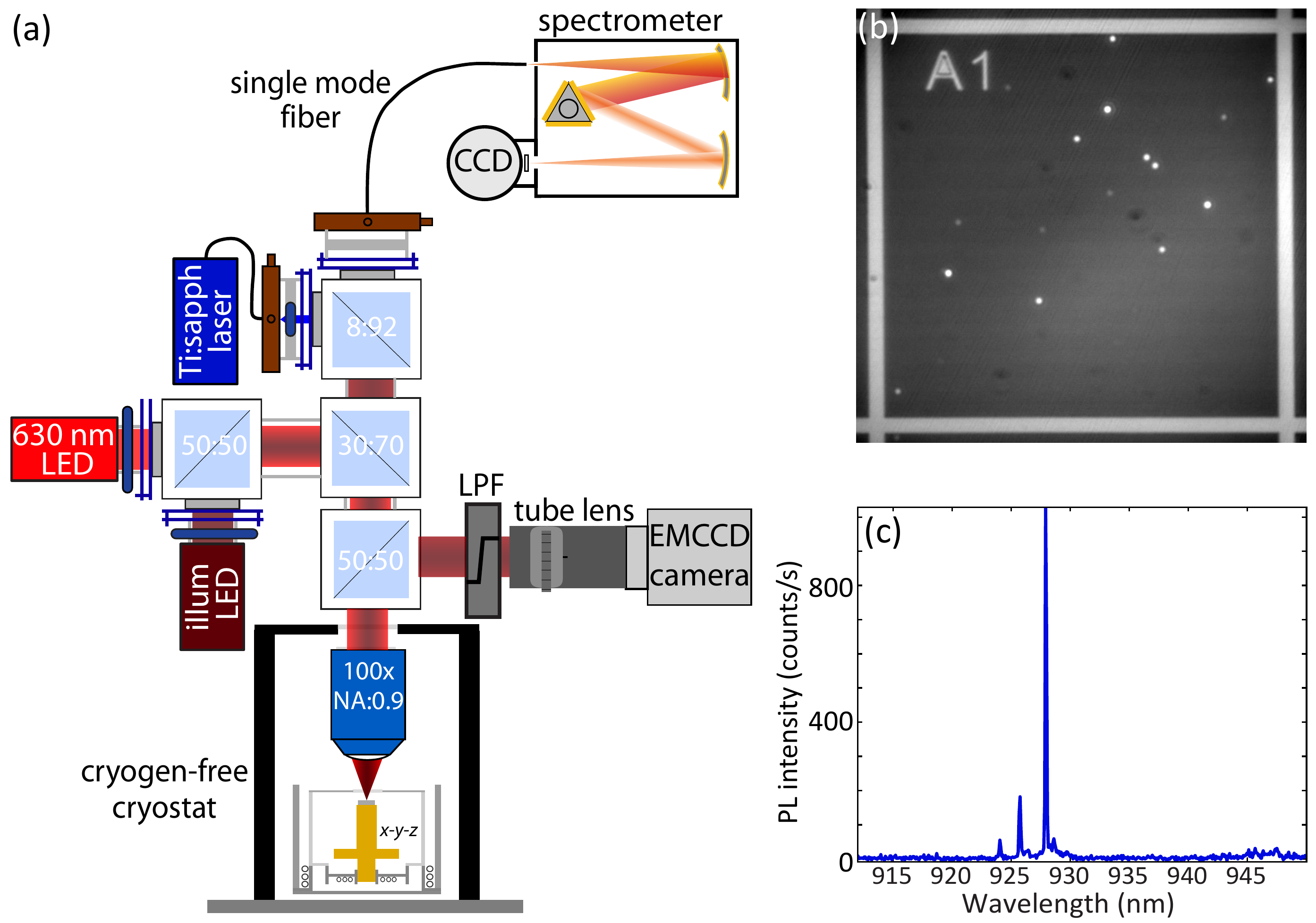}
   \caption{(a) Schematic of the cryogenic photoluminescence imaging system.  The sample sits on an $x$-$y$-$z$ positioning stack within a 4~K closed-cycle cryostat.  Hanging above the sample, within the cryostat, is a 100$\times$ magnification, 0.9 NA microscope objective.  Two different color LEDs are combined on a 50:50 beamsplitter and sent into the objective in order to excite the quantum dots (QDs) and illuminate the sample, respectively.  The QD excitation LED is typically chosen as 630~nm, while the illumination LED depends on the specifics of the material epitaxy, in particular, the underlying layers, and is at 940~nm.  Emitted light from the QDs and reflected light from the sample are either directed through one or more long pass filter (LPFs) to reject the quantum dot excitation light and wetting layer emission before going into a tube lens and an EMCCD camera, or are coupled into a single mode fiber and sent to a grating spectrometer for spectral analysis.  In addition, a tunable wavelength, continuous-wave Ti:sapphire laser is coupled into the system from single mode fiber through an 8:92 beamsplitter, to enable excitation of a small number of QDs within a focused laser spot. (b) Representative image acquired by the EMCCD camera in a 1~s integration time. (c) Representative photoluminescence spectrum from a single QD.}
   \label{fig1}
\end{figure}

Our positioning approach is based on wide-field imaging in which two LED sources are used, one for excitation of the QDs and the other for illumination of the sample (see Fig.~\ref{fig1}(a)).  The light emitted by two LEDs is combined on a 50:50 beamsplitter and sent through two additional beamsplitters above the cryostat (30:70 and 50:50) before reaching the cryostat objective.  The light emitted from the semiconductor material and reflected off its top surface consists of many different colors, namely reflected QD excitation light (630~nm), QD emission (900~nm to 950~nm range), emission from other recombination centers in the material (typically <860~nm, and originating from the GaAs band edge, carbon acceptor states, and wetting layer states), and reflected illumination (typically 940~nm).  The 50:50 beamsplitter above the cryostat sends the reflected and emitted light towards an EMCCD (electron multiplying charge coupled device) camera in the reflected path, and through the 30:70 beamsplitter and an 8:92 beamsplitter in the transmitted path. By placing appropriate filters (notch filters and/or edge-pass filters) in front of the EMCCD camera to eliminate all light other than emission from the QD states and the reflected illumination, we obtain images such as those shown in Fig.~\ref{fig1}(b).  The EMCCD camera chip is 1004 pixels~$\times$~1002 pixels, with each pixel having dimensions of 8~$\mu$m~$\times$~8~$\mu$m.  The 8:92 beamsplitter allows for introduction of pump light from a continuous-wave Ti:sapphire laser and for collection of light emitted from the sample into a single mode fiber, where it can be sent to a grating spectrometer for spectral analysis.  The focused pump spot from the Ti:sapphire laser is $\lesssim2$~$\mu$m in diameter, and allows for selection of individual QDs from within the 66~$\mu$m~$\times$~66~$\mu$m field of view provided by the imaging system (the tube lens focal length of 200~mm is larger than the nominal 165~mm for the 100$\times$ infinity-corrected objective, resulting in an additional magnification of 1.2$\times$ before the EMCDD).  An example photoluminescence spectrum from an individual QD is shown in Fig.~\ref{fig1}(c).

\begin{figure*}[t]
  \includegraphics[width=0.7\linewidth]{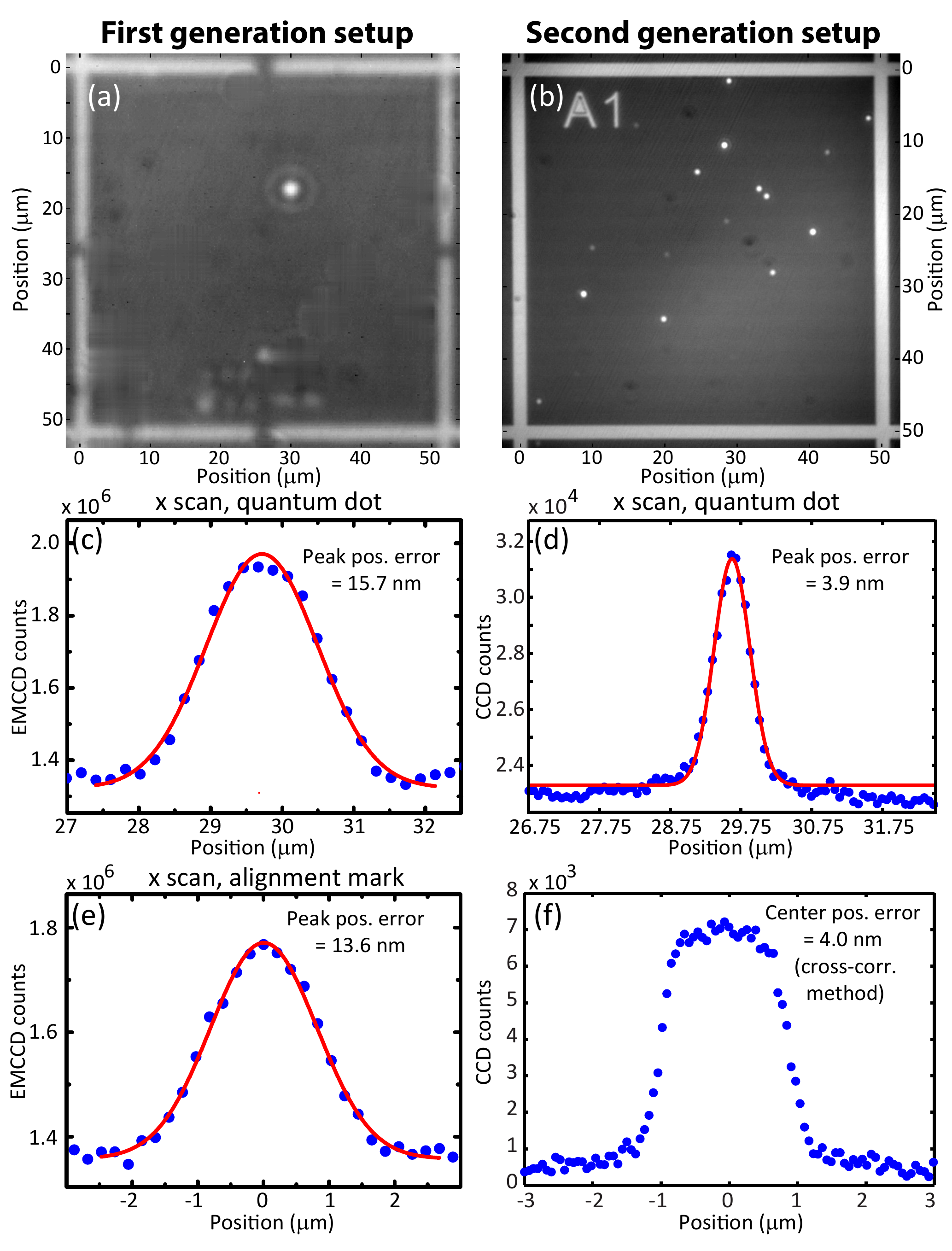}
   \caption{Comparison of the images collected with the system developed in Ref.~\onlinecite{sapienza_nanoscale_2015} (left column panels (a), (c), and (e): First generation setup) and the system described in this work (right column panels (b), (d), and (f): Second generation setup). (a)-(b) Images of the QD photoluminescence and alignment marks. (c)-(d) One-dimensional line cut of the QD emission (along the $x$-axis). (e)-(f) One-dimensional line cut of the light reflected off an alignment mark (along the $x$-axis).  The uncertainty in the peak position of the QD emission (Gaussian fit in(c); maximum likelihood estimate in (d)) and the center of the alignment mark (Gaussian fit in (e), cross-correlation method in (f)) are stated on the graphs, and represent one standard deviation values and 68~$\%$ confidence intervals, respectively, as detailed in Section~\ref{sec:Analysis_Methods}.}
\label{fig2}
\end{figure*}

\begin{figure*}[t]
  \includegraphics[width=0.7\linewidth]{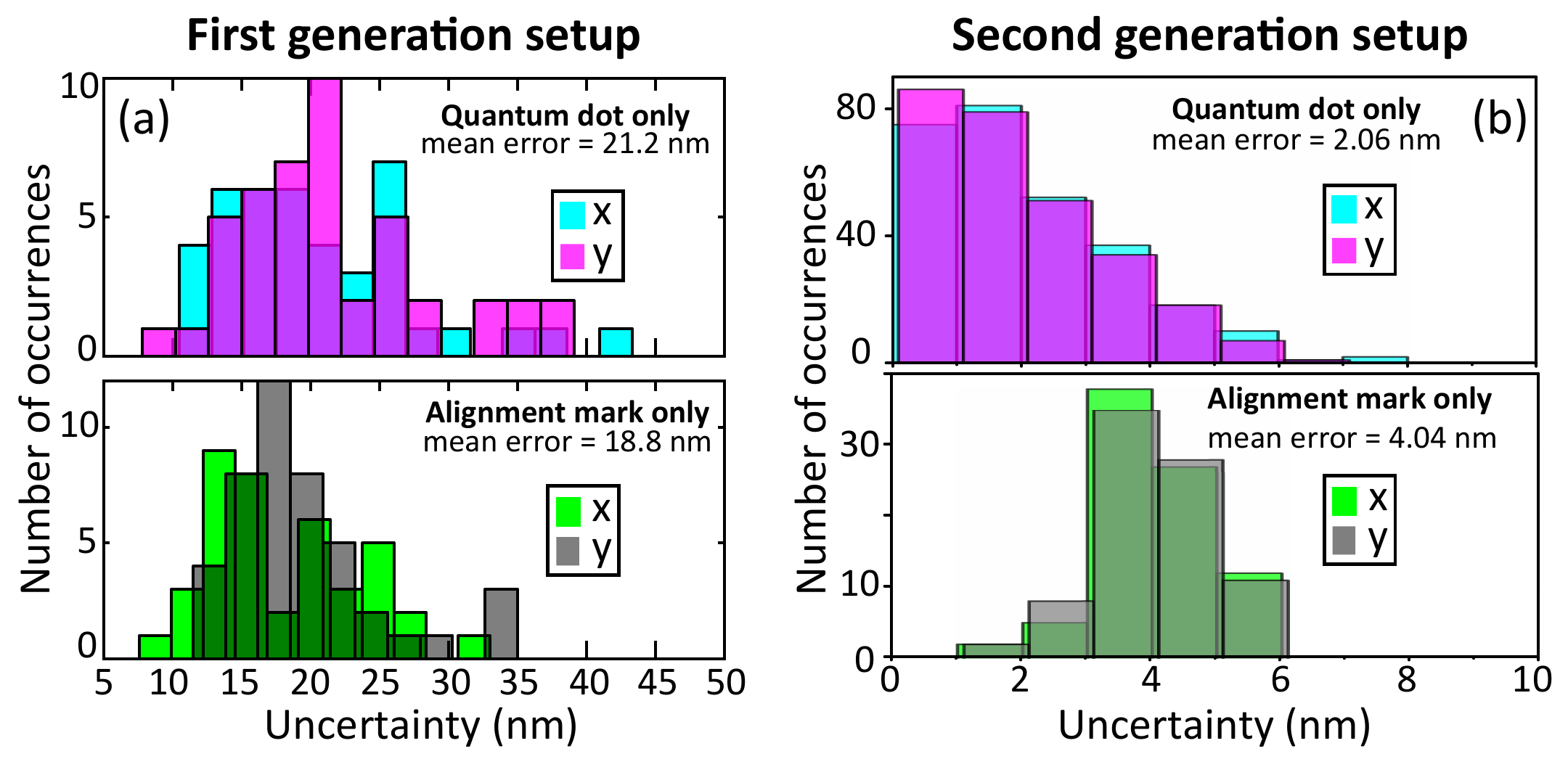}
   \caption{Histograms of the uncertainties in the QD and alignment mark positions for the (a) first generation setup and (b) system described in this work (second generation setup).  Compared to the first generation setup, the mean QD position uncertainty in the second generation setup is reduced by a factor of 10.3, while the mean alignment mark position uncertainty is reduced by a factor of 4.7.  The mean uncertainty of the QD-alignment mark separation is $\approx$~28~nm for the first generation setup and $\approx4.5$~nm for the second generation setup.  The methods used for uncertainty estimation are detailed in Section~\ref{sec:Analysis_Methods}.}
   \label{fig3}
\end{figure*}

The system is constructed using cage-mounted commercial optical and optomechanical components, with the exception of a central aluminum block (not shown) that provides structural rigidity. The use of an objective within the cryostat provides several advantages and tradeoffs in comparison to setups in which the objective is held outside of the cryostat, as in Ref.~\onlinecite{sapienza_nanoscale_2015}, in which a long working distance 20$\times$ objective (0.4 NA) was used.  The geometry of the cryostat is such that the vibrations induced by the cryocooler, which in our case is based on a Gifford-McMahon architecture, are common to both the objective and the sample.  The ability to place the objective in close proximity to the sample allows for the use of a high NA (0.9), which both increases the solid angle over which emitted photons are collected, and also increases the LED intensity at the sample, ensuring that saturation of all QDs within the field of view can be achieved (this was not the case in Ref.~\onlinecite{sapienza_nanoscale_2015}).  The net result is to increase the collected photon flux from the QD, thereby reducing the camera gain (no EM gain is needed) and generally improving the measurement dynamic range and signal-to-noise ratio, and reducing the required image acquisition time. Finally, the absence of optical windows between the objective and the sample leads to higher quality imaging (for example, alignment features appear sharper), as we discuss below.  On the other hand, the larger magnification, combined with the fixed size of the EMCCD camera chip, results in a smaller field of view, and the limited translation range of the nanopositioners within the cryostat limits the sample area that can be covered in comparison to the previous work, where the entire microscope system could be moved independently from the sample and cryostat.

Figure~\ref{fig2} presents a comparison of the results from the imaging setup described in ref.~\onlinecite{sapienza_nanoscale_2015} (First generation setup) and the new system (Second generation setup), where the samples investigated emit in the 930~nm wavelength band.  The modified system results in several readily observable differences in the collected images.  First, the alignment marks in the image from the second generation setup (Fig.~\ref{fig2}(b)) are noticeably sharper than those from the first generation setup (Fig.~\ref{fig2}(a)), a result of both the higher resolution of the imaging system and more importantly, the absence of image-distorting optical windows between the objective and the sample.  This is seen quite clearly when considering line scans across an alignment mark (Fig.~\ref{fig2}(e)-(f)), where the blurring produced by the windows results in an essentially Gaussian shape for the line scan in the first generation setup, while the line scan for the second generation setup, though not perfectly rectangular, show much more abrupt edges.  The higher spatial resolution provided by the larger magnification (increased from a 40$\times$ system magnification to a 100$~\times$ system magnification) results in the full-width at half-maximum of the QD emission being reduced from $\approx$~2~$\mu$m to $\approx$~650~nm, which can be important in order to clearly distinguish between QDs when working with higher QD density samples.  The improved signal-to-noise level that results from increased collected photon flux results in significantly reduced fit uncertainties in the center positions of the QD emission and alignment marks (the specific procedures used are detailed in Section~\ref{sec:Analysis_Methods}), and moreover, the acquisition time required to achieve such uncertainties is only 1~s, in comparison to 120~s for the first generation setup.  This reduction significantly increases the throughput, limits the potential influence of drift during the measurements, and stands in contrast to serial detection techniques, such as scanning confocal microscopy, which require a similar integration time (assuming equal collection efficiency) for each pixel in the image~\cite{thon_strong_2009}.

\begin{table*}[t]
\centering
\vspace{0.1cm}
\begin{threeparttable}
\caption{Comparison of the performance of the PL imaging setup from Ref.~\onlinecite{sapienza_nanoscale_2015} (First generation setup) and the current work (Second generation setup).}
\label{Table1}
\begin{tabular}{|c||c|c|}
\hline
\vspace*{\fill}  &   \textbf{First generation setup} &  \vspace*{\fill}  \textbf{Second generation setup} \\
 \hhline{|=||=|=|}
  \textbf{Objective location}   & Outside cryostat & Inside cryostat \\
 \hline
 \textbf{Objective NA}   & 0.4    & 0.9 \\
 \hline
  \textbf{Objective magnification} & 20$\times$ & 100$\times$ \\
 \hline
  \textbf{System magnification} & 40$\times$ & 120$\times$ \\
 \hline
 \textbf{Usable field of view} & 200~$\mu$m~$\times$~200~$\mu$m & 66~$\mu$m~$\times$~66~$\mu$m\tnote{a} \\
 \hline
 \textbf{EM gain} & 200 (typical) & None \\
 \hline
\textbf{Integration time} & 120~s & $<1$~s \\
\hline
\textbf{Mean QD position uncertainty} & 21~nm\tnote{b} &  2~nm\tnote{c}\\
\hline
\textbf{Mean alignment mark position uncertainty} & 19~nm\tnote{d} & 4~nm\tnote{e}\\
\hline
 \end{tabular}
 \begin{tablenotes}
\item[a] Using a 165~mm focal length tube lens will result in an 80~$\mu$m~$\times$~80~$\mu$m field of view.
\item[b] One standard deviation value from one-dimensional Gaussian fit.
\item[c] One standard deviation value from a two-dimensional maximum likelihood estimate (see Section~\ref{sec:Analysis_Methods}).
\item[d] One standard deviation value from one-dimensional Gaussian fit.
\item[e] 68~$\%$ confidence interval from a cross-correlation method (see Section~\ref{sec:Analysis_Methods}).
\end{tablenotes}
\end{threeparttable}
\end{table*}

Figure~\ref{fig3} shows a comparison of the fit uncertainties for the QDs and alignment marks in the two different setups, along both the $x$ and $y$ directions, for a number of interrogated QDs. The mean uncertainty in the QD emission center is reduced by a factor of 10.3, while the uncertainty in the alignment mark center is reduced by a factor of 4.7.  No significant difference is seen between the $x$ and $y$ directions in either setup.  A summary comparison of the two setups is provided in Table~\ref{Table1}.  Finally, we note that while photoluminescence imaging and basic non-resonantly-pumped spectroscopy of the quantum dots is the main focus of this work, we have been able to make simple modifications to the second generation setup to enable a wider variety of measurements to be performed.  For example, the insertion of polarization-controlling optics in the Ti:sapphire laser excitation path and crossed polarizers before the objective and before the collection fiber have been successfully used to observe resonance fluorescence~\cite{ref:Warburton_res_fluor} from single QDs.  In addition, the collection efficiency of the setup is adequate to perform high-resolution spectroscopy (e.g., with a scanning Fabry Perot analyzer) and photon correlation measurements of single QDs in bulk.

\section{Image Analysis Methods}
\label{sec:Analysis_Methods}

\begin{figure*}[h]
  \includegraphics[width=0.75\linewidth]{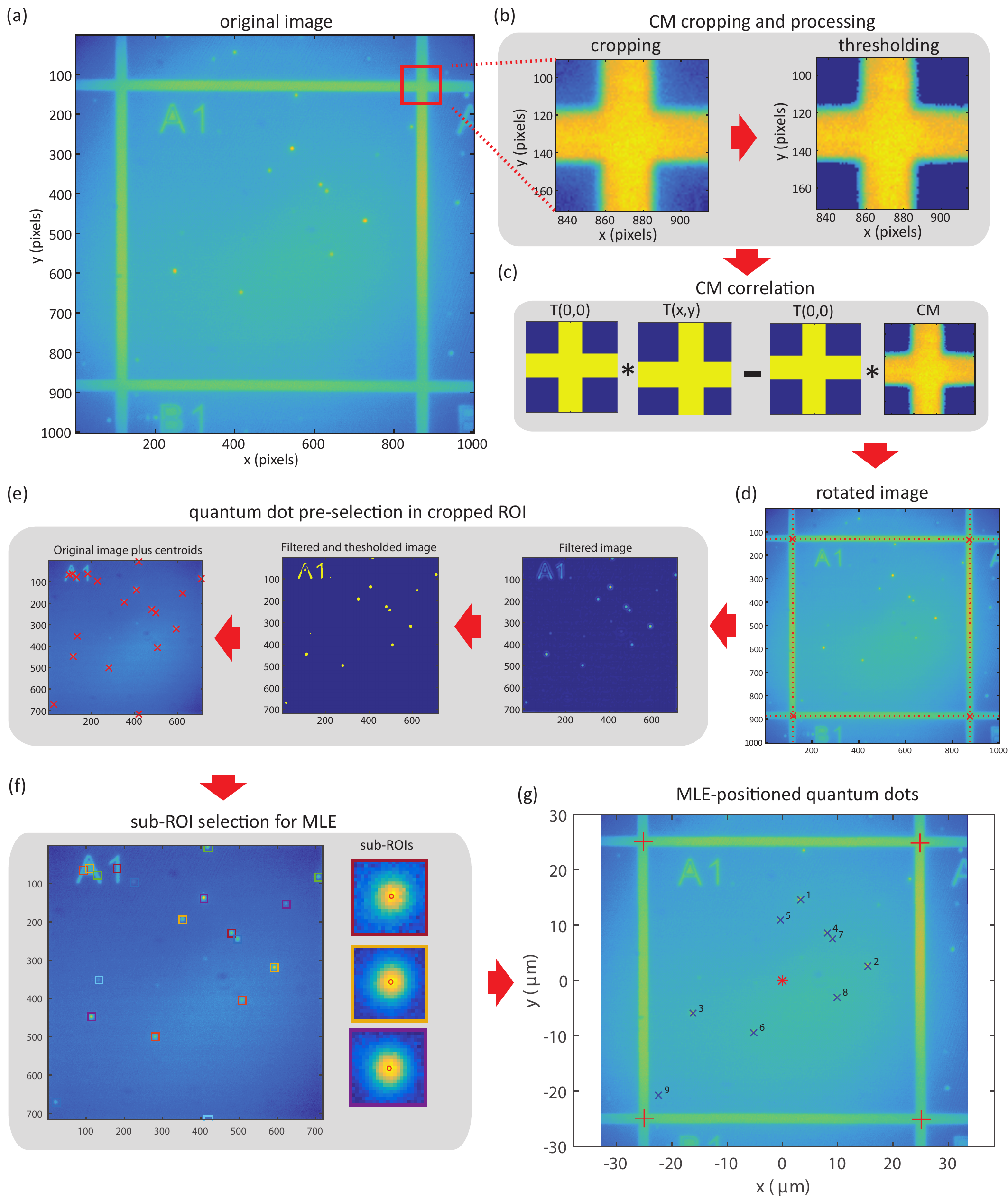}
   \caption{Summary of the process by which the photoluminescence images are analyzed. (a) Original image. (b) Sections of the original image near the four alignment marks are cropped and filtered through threshold detection. (c) The alignment mark locations in the original image are determined through an approach in which the convolution of an ideal alignment mark pattern (T(0,0)) and the processed alignment mark pattern (CM) is compared with the autocorrelation of the ideal alignment mark pattern, and the residual is minimized. (d) The location of the four alignment marks is used to rotate the image so that it is better aligned with respect to the Cartesian $x$-$y$ axes. This process of alignment mark correlation and image rotation is iterated until the residual stabilizes, and then a final rotated image is produced. (e) Pre-selection of quantum dots based on filtering, thresholding, and centroiding. (f) Definition of sub regions of interest (ROI) within the space defined by the chip marks.  A two-dimensional Gaussian maximum likelihood estimate (MLE) is used to localize the emission centers of each QD within its sub-ROI. The insets show zoom-in images of three different QDs, with the red circles indicating the MLE determination of the centers. (g) The alignment mark data and QD data are displayed in a final analyzed image, with the alignment mark centers given by red + signs, the center of the field given by a red $\ast$ sign, and the QDs (labeled 1-9) given by blue x signs.}
   \label{fig4}
\end{figure*}

In this section we review the techniques used to analyze the photoluminescence images and determine the uncertainties when localizing the QD emission and the alignment mark center. In Ref.~\onlinecite{sapienza_nanoscale_2015}, one-dimensional nonlinear least squares fits to Gaussian functions along the $x$-axis and $y$-axis were used, after correcting for any rotation in the image, and the reported uncertainties were one standard deviation values based on the fits. Similar approaches have been used in other QD positioning studies, for example, those based on scanning confocal photoluminescence~\cite{lee_registration_2006,dousse_controlled_2008,thon_strong_2009}.  Here, we have moved beyond this simple approach to take into account: (1) the sharpness of the alignment marks, which in ref.~\onlinecite{sapienza_nanoscale_2015} could be adequately fit by Gaussians due to the blurring caused by the cryostat windows, and (2) advances in single-molecule localization~\cite{thompson_precise_2002,smith_fast_2010,mortensen_optimized_2010}, based on the use of a maximum likelihood estimator (MLE) as a means to approach the theoretical minimum uncertainty in positioning a fluorophore given the available information (e.g., number of photons in the image and the system's point spread function (PSF)).

Our basic image analysis approach is summarized in Fig.~\ref{fig4}. We use two different image processing techniques to obtain the spatial location of both the four alignment marks and the QDs with subpixel accuracy. To locate the alignment marks, we employ an algorithm developed for use in direct-write electron beam lithography~\cite{anderson_sub-pixel_2004}. For each of the four marks, a corresponding section of the image (Fig.~\ref{fig4}(a)) is cropped, flattened, and thresholded, to yield a processed chipmark image, denoted by CM (Fig.~\ref{fig4}(b)). We next calculate the root-mean-square value of the residual T$(x_i,y_i)\ast $CM - T$(0,0)\ast$T$(x_i,y_i)$ across all pixels, where T$(x_i,y_i)$ is a chipmark template centered at pixel $(x_i,y_i)$ and $\ast$ denotes a 2D convolution. This process, illustrated in Fig.~\ref{fig4}(c), is repeated for every $(x_i,y_i)$ pixel in CM, producing a 2D residual map R$(x_i,y_i)$ that is then fitted with a two-variable quadratic function $P(x,y) = a_0+a_1x a_2+x^2+a_3xy + a_4y^2 + a_5xy$, where $x$ and $y$ are in pixel units. The point $(x,y)_\text{min}$ that minimizes $P$ corresponds to the center of the CM, and can be obtained by solving $\partial P/\partial x=\partial P/\partial y=0$ and substituting the fitting coefficient values. Uncertainties in the alignment mark position are obtained by propagating the 68$~\%$ parameter confidence intervals from the fit (68~$\%$ confidence intervals were chosen because they provide the most direct comparison to the one standard deviation uncertainties obtained in Gaussian fits to determine alignment mark locations in Ref.~\onlinecite{sapienza_nanoscale_2015}). Importantly, the width of the arms of the cross forming the template pattern must be chosen so as to maximize the contrast in the R$(x_i,y_i)$, and thus the correlation between the experimental image and the template. After the centers of the four alignment marks are found though this procedure, a rotation correction is applied to ensure that the alignment marks are aligned with respect to the Cartesian $(x,y)$ axes (Fig.~\ref{fig4}(d)). This process is repeated until the rotation angle converges to a desired tolerance. At this point, a pixel size in real space units can be determined by comparing the pixel distances between the positioned mark centers and their nominal, real space values. All coordinates in the figure are then referenced to the centroid of the four positioned mark locations, which is the write field center.

The next step consists of using a MLE to position the imaged QD emission with respect to the write field center. In the PL image, a region of interest (ROI) is defined to comprise the area bounded by the four alignment marks. Bright spots due to the QD's emission within the ROI image are first pre-selected as follows (Fig.~\ref{fig4}(e)). The image is first filtered with a low-pass followed by a (complementary) high-pass Gaussian filter with adjustable variance, which enhances the image signal-to-noise ratio. The image is then thresholded and the QD spots are localized by centroiding. These centroids are then used as center coordinates for sub-ROIs comprising a small number of user-specified pixels on the unfiltered image (colored boxes in Fig.~\ref{fig4}(f)), to which the MLE algorithm is next applied. Our MLE algorithm is based on a 2D Gaussian PSF for the QDs, as described in Ref.~\onlinecite{mortensen_optimized_2010}, and is briefly described below.

At each sub-ROI, $\left\{Y_k\right\}_{k=1}^{n}$ is a sequence of independent random variables that represents the number of detected photon counts in the k-th pixel. We assume $Y_k$ follows Poisson statistics, with an expected number of photon counts $\lambda_k$ at pixel $k$. We then let $\lambda_k = \lambda_k(\theta)$, where $\theta \in \Theta$, and $\Theta$ is an appropriate parameter set that depends on the point spread function model. As noted above, our PSF model consists of a symmetric 2D Gaussian with variance $\sigma^2$, centered at $(i_k,j_k)$. The probability mass function (PMF) for $Y_k$ is
\begin{equation}
f_{\theta}(y_k) = \dfrac{{\lambda_k^{y_k}(\theta)}}{y_k!}\exp\left[-\lambda_k(\theta)\right], y_k \in \mathbb{N}
\end{equation}
where
\begin{equation}
\lambda_k(\theta) = \alpha\cdot p(i_k,j_k) + \beta,
\end{equation}
$\theta = (i_0,j_0,\sigma^2,\alpha,\beta)^{\mathrm{T}}$ is the parameter vector, $\alpha$ is the total expected  number of photon counts, $p$ is Gaussian PSF, and $\beta$ is the expected background number of photon counts per pixel. The joint PMF among the number of photon counts of $n$ pixels is $f^{(n)}_{\theta}(y^n) = \prod_{k=1}^{n} f_{\theta}(y_k)
$. The MLE for every $\lambda_k$ is $g_\text{MLE} (y^n) = \argmax_{\theta \in \Theta}\left\{f^{(n)}_{\theta}(y^n)\right\}$,
which can be written as
\begin{align}
g_\text{MLE} (y^n) = \argmin_{\theta \in \Theta}\left\{\sum_{k=1}^{n}\lambda_k(\theta) - y_k\ln\lambda_k(\theta)\right\}.
\label{eq:MLE}
\end{align}
The parameters in $\theta$ are estimated by minimizing the expression inside curly brackets in eq.~\ref{eq:MLE}. We employ the Nelder-Mead simplex method for this purpose.

Uncertainties for the QD position (as well as $\sigma^2$, $\alpha$ and $\beta$) can be obtained by calculating the Cram{\'e}r-Rao lower bounds, which are the asymptotic variances for the estimated parameters (i.e., the MLE variances monotonically decrease to the Cramer-Rao lower bounds as the number of photon counts goes to infinity). These bounds are obtained from the diagonal elements of the inverse of the Fisher Information matrix $M(\theta)$,
\begin{equation}
M(\theta) \triangleq \mathbb{E}_\theta\left[\nabla_\theta\textrm{ln}f^{(n)}_\theta(Y^n)\left(\nabla_\theta\textrm{ln}f^{(n)}_\theta(Y^n) \right)^{\textrm{T}}  \right],
\end{equation}
whose elements can be shown to be
\begin{equation}
\left[M(\theta)\right]_{rs} = \sum_{k=1}^{n}\dfrac{\partial \lambda_k(\theta)}{\partial \theta_s}\dfrac{\partial \lambda_k(\theta)}{\partial \theta_r}\dfrac{1}{\lambda_k(\theta)}
\end{equation}
and calculated analytically. Typical uncertainty values are ($\delta$x$_\text{MLE}$ and $\delta$y$_\text{MLE}$ with values inside square brackets in Table~\ref{Table2}),
shown in Fig.~\ref{fig3}, and correspond to one standard deviation for the Gaussian center. It is important to note that the electron multiplication mechanism of EMCCDs leads to non-Poissonian statistics for the detected signal, which should be taken into account in a strict MLE uncertainty analysis~\cite{mortensen_optimized_2010,hirsch_stochastic_2013}. In practice, however, the main effect of the multiplication process is that it adds excess noise that leads to a position standard deviation (i.e., localization uncertainty) larger by a factor of $\sqrt{2}$.  Moreover, in our current setup, we typically do not use EM gain (Table~\ref{Table1}). The final result produced by our analysis routine consists of an image with the alignment mark and QD positions labeled (Fig.~\ref{fig4}(g)) and a table of positions of the QDs with respect to the center of the field, including the chip mark center uncertainties (Table~\ref{Table2}), 
that can be used in subsequent device fabrication, including aligned electron-beam lithography.

\begin{table}[t]
\caption{Quantum Dot Positions and Uncertainties ($\delta$x and $\delta$x$_\text{MLE}$ are the uncertainties for QD-alignment mark separation and QD position respectively)} 
\centering 
\begin{tabular}{c c c c c} 
\hline\hline 
QD~\# & x ($\mu\text{m}$) & y ($\mu\text{m}$) & $\delta$x [$\delta$x$_\text{MLE}$] (nm) & $\delta$y [$\delta$y$_\text{MLE}$](nm)\\ [0.5ex]
\hline 
1 & 3.34 & 14.60 & 4.6 [0.81] & 4.4 [0.81]\\
2 & 15.47 & 2.56 & 4.6 [1.06] & 4.5 [1.06]\\
3 & -16.16 & -5.83 & 4.6 [1.10] & 4.5 [1.10]\\
4 & 8.08 & 8.56 & 4.7 [1.37] & 4.6 [1.37]\\
5 & -0.41 & 10.92 & 4.8 [1.60] & 4.7 [1.60]\\
6 & -5.15 & -9.35 & 4.8 [1.74]& 4.7 [1.74]\\
7 & 9.11 & 7.52 & 4.8 [1.73]& 4.7 [1.73] \\
8 & 9.89 & -3.0 & 5.0 [2.22] & 4.9 [2.22]\\
9 & -22.34 & -20.68 & 5.2 [2.60] & 5.1 [2.60]\\ [1ex]
\hline 
\end{tabular}
\label{Table2} 
\end{table}

\section{Conclusions}

In summary, we have presented an improved implementation of the two-color photoluminescence imaging approach used in Ref.~\onlinecite{sapienza_nanoscale_2015} to locate the position of single self-assembled QDs with respect to alignment marks for subsequent fabrication with single QDs optimally aligned within nanophotonic devices.  Our new approach reduces the uncertainties in the QD positions by a factor of 6, and the image acquisition time needed to achieve those uncertainties by a factor of 100, yielding a mean one standard deviation uncertainty in the QD location of 4.5~nm. The superior optical performance and more sophisticated and automated image analysis tools have greatly increased the throughput of this technique. As a result we believe that it will be valuable in future demonstrations of QD-based devices for integrated quantum photonics~\cite{dietrich_gaas_2016}.  In the future, a combination of automated sample positioning, focus adjustment (through the sample height), and stage encoder readout may enable stitching of multiple fields together, further increasing the throughput and capability of the system.  Moreover, we anticipate that this technique can be easily extended to study other solid-state quantum emitters, including in emerging systems such as SiC~\cite{castelletto_silicon_2013} or two-dimensional semiconductors~\cite{perebeinos_metal_2015}.

\vspace{0.25cm}
\section*{Acknowledgements}
JL acknowledges support under the Cooperative Research Agreement between the University of Maryland and NIST-CNST, Award 70NANB10H193, National Natural Science Foundation of China (grant no.~11304102) and the Ministry of Science and Technology of China (grant no.~2016YFA0301300). J.V.D.M.C. acknowledges funding from the Brazilian Ministry of Education through the Brazilian Scientific Mobility Program CAPES-grant 88888.037310/2013-00. JDS acknowledges supports from the KIST Internal Program of flag-ship (2E26420).  The authors thank J.A. Liddle from NIST for introduction to the cross-correlation method for alignment mark detection, and Kerry Neal and Caleb Schreibeis from Montana Instruments for assistance with the cryostat system.

\end{document}